\documentstyle[preprint,tighten,aps]{revtex}
\begin{document}
\draft
\newcommand{\lmx}[1]{\begin{displaymath} {#1}=
                    \left(\begin{array}{rrrr}}
\newcommand{\rmx}{\end{array} \right) \end{displaymath}}
\title{\bf Low-temperature renormalization group study
of uniformly frustrated models for type-II superconductors }
\author{ Giancarlo Jug }
\address{
Max-Planck Institut f{\"u}r Physik komplexer Systeme,
Au{\ss}enstelle Stuttgart, Heisenbergstr. 1, Postfach 800665
D-70569 Stuttgart (Germany) \cite {GJ} }
\author{ Boris N. Shalaev }
\address{ A.F.Ioffe Physical \& Technical Institute, Russian Academy
of Sciences, \\
194021 St.Petersburg (Russia) \cite{BNS} \\
and Max-Planck Institut f{\"u}r Physik komplexer Systeme, D-70569
Stuttgart (Germany)}

\maketitle

\begin{abstract}
\noindent
We study phase transitions in uniformly frustrated $SU(N)$-symmetric
$(2+\epsilon)$-dimensional lattice models describing type-II 
superconductors near the upper critical magnetic field $H_{c2}(T)$. The 
low-temperature renormalization-group approach is employed for calculating 
the beta-function $\beta(T,f)$ with $f$ an arbitrary rational magnetic
frustration. The phase-boundary line $H_{c2}(T)$ is the ultraviolet-stable
fixed point found from the equation $\beta(T,f)=0$, the corresponding 
critical exponents being identical to those of the non-frustrated continuum 
system. The critical properties of the $SU(N)$-symmetric complex
Ginzburg-Landau (GL) model are then examined in $(4+\epsilon)$ dimensions.
The possibility of a continuous phase transition into the mixed state
in such a model is suggested.
\end{abstract}

\vfill
\newpage

\section{ INTRODUCTION }
\renewcommand{\theequation}{1.\arabic{equation}}
\setcounter{equation}{0}

The challenging problem of the critical behavior of a type-II 
superconductor near the upper critical magnetic field $H_{c2}(T)$ has a 
rich and long history going back to the seminal paper by Eilenberger 
\cite{a1} (see also \cite{a2}).
It has been recognized for some time that an external magnetic field 
drastically changes the critical properties of superconductors. The 
magnetic field hinders the growth of the thermal fluctuations in the plane 
perpendicular to ${\bf H}$, since the growth of their correlation length
is restricted by the magnetic length scale $\ell=\sqrt{hc/eH}$ which is
much shorter than the coherence length $\xi$. This effect of dimensional
reduction results in an enhancement of the longitudinal fluctuations 
leading, in particular, to the increase of the lower critical dimension 
from 2 to 4.

If critical fluctuations are ignored, the uniform frustration (in the
language of spin models of the vortex lattice) eventually leads to a 
continuous phase transition into the Abrikosov flux lattice state.
On the one hand, in contrast to mean-field theory, the standard
renormalization group (RG) approach in $6-\epsilon$ dimensions, in fact, 
failed so far to yield insight on the nature of the phase transition due 
to the appearence of an infinite number of invariant charges (relevant 
scaling variables) inherent to the non-renormalizable scalar $\phi^{4}$ 
field theory in a field \cite{a3}. 
Physical arguments (presented e.g. in \cite{a2}) support the existence of 
a first-order melting transition for the flux lattice. On the other hand, 
however, the conventional $1/N$-expansion, when applied to the 
$SU(N)$-symmetric Ginzburg-Landau (GL) model with an $N$-component order 
parameter, gives a second-order phase transition above four dimensions 
\cite{a4,a5,a6}. The question of the nature of the phase transition from 
the normal into the mixed state of a type-II superconductor remains 
therefore an open problem, even before the effects of impurities and the 
topical question of the vortex glass are to be considered. In this paper 
we wish to investigate what is to some extent a new question, namely the 
nature of this phase transition when a periodic lattice potential is 
coupled to the superconducting order parameter.

It was shown some time ago  \cite{a7,a8} that the interaction between 
thermal fluctuations and the underlying crystal lattice can restore the 
phase transition into the mixed state. It is well known that the vortex 
lattice state has two broken symmetries: i) phase coherence, that is
off-diagonal long-range order (ODLRO), and ii) translational symmetry. 
Including the interaction with an underlying lattice means that the 
translational symmetry is explicitly broken and lattice effects (for 
instance Harper's broadening and splitting of the Landau levels suppressing
infinite degeneracy of the energy spectrum) become particular acute.

Lattice models of superconductors were introduced long ago and are of
great interest by their own accord, being employed to study phase 
transitions in uniformly frustrated XY models \cite{a9}, or in some 
artificial condensed matter structures like two-dimensional (2D) 
Josephson-junction arrays subject to a magnetic field \cite{a10} or in
opals filled with a semiconductor or a metal \cite{a101}.

Naturally, according to a naive point of view, near criticality the 
correlation length $\xi$ diverges and the system should ``forget'' about 
the discretness of the underlying lattice; however, this turns out not to 
be the case. In fact, in the continuum limit some specific uniformly-
frustrated lattice models have been shown not to be equivalent to the 
conventional GL model \cite {a9}. If lattice effects are taken into account, 
in the long-wavelength limit one obtains an infinite set of effective 
(renormalizable) GL Hamiltonians corresponding to different rational values 
of the frustration $f=\Phi/\Phi_{0}$, where $\Phi=Ba^{2}$ is the magnetic 
flux per plaquette and $\Phi_{0}=hc/2e$ the elementary flux quantum 
\cite{a5,a9}.

The main goal of the present paper is to consider the critical behavior of 
lattice models of type-II superconductors in the normal phase above
$H_{c2}(T)$. We shall apply the standard low-temperature RG approach that 
was first introduced and developed in \cite{a11,a12}. The method 
was first exploited for studying critical phenomena in superconductors 
(with $B=0$) in \cite{a17}. We will show how the presence of a lattice 
induces a second-order phase transition in dimensions $d=2+\epsilon$, by 
giving explicitly the phase boundary in the low-temperature limit.

The remainder of this paper is organized as follows. In Section II the 
critical behavior of the $(2+\epsilon)$-dimensional $SU(N)$-symmetric 
lattice model is considered by means of the low-temperature RG approach. 
Section III deals with the standard GL model in $4+\epsilon$ dimensions, 
therefore in the absence of the underlying lattice, and we show that the 
phase transition appears to be again of second order, although it remains 
impossible to reach a conclusion about the nature of the phase transition 
in $d=3$. Section IV contains a discussion of our results and of related 
issues in which lattice effects may play an important role and some 
concluding remarks.

\section{ VORTEX-LATTICE PHASE TRANSITION IN $2+\epsilon$ DIMENSIONS }
\renewcommand{\theequation}{2.\arabic{equation}}
\setcounter{equation}{0}

\subsection{Effective action and low-temperature expansion}

We begin by considering the classical Hamiltonian of the $SU(N)$-symmetric
nonlinear sigma model, defined on a square lattice with periodic boundary 
conditions:

\begin{equation}
H=-J \sum_{<i,j>}|S_{i}^{a}\exp(iA_{ij})-S_{j}^{a}|^{2}
\label{e1}
\end{equation}

\noindent
This is indeed the lattice version of the GL Hamiltonian, $ S_{i}^{a} $ 
being a $N$-component complex unit vector with the constraint 
$ S_{i}^{a}S_{i}^{a*}=1 $, $a=1,2,...,N$ and $J$ is a coupling constant.
Here $<\cdots>$ indicates that the summation is over all nearest-neighboring
sites, as usual. $A_{ij}$ is a bond angle such that the sum around a
plaquette is given by

\begin{equation}
\sum_{plaquette}A_{ij}=2{\pi}f
\label{e2}
\end{equation}

\noindent
with $f=p/q$ the so-called frustration; $p$ and $q$ are here mutually prime
integers and $A_{ij}$ is defined by

\begin{equation}
A_{ij}=\frac{2\pi}{\Phi_{0}} \int_{i}^{j}dx_{\mu}A_{\mu}
\label{e3}
\end{equation}

\noindent
where $A_{\mu}=(-By,0,0)$ is the vector potential of the uniform
magnetic field $B$ along the $z$ axis. From eq.(\ref{e2}) it follows that 
the magnetic flux through a plaquette is assumed to be a rational fraction 
of the magnetic flux quantum $\Phi_{0}$.

Consider the partition function associated with eq.(\ref{e1}), namely

\begin{equation}
Z=\int \prod_{i}d^{N}S_{i}d^{N}S_{i}^{*}\exp\{-\frac{H}{T}\}
\delta( S_{i}^{a}S_{i}^{a*}-1)
\label{e4}
\end{equation}

\noindent
Notice that the Hamiltonian in eq.(\ref{e1}) has both a global nonabelian
$SU(N)$ symmetry and a local $U(1)$ one whilst the fixed-length constraint 
imposed on the local spins is $0(2N)$-symmetric.

The main steps in our calculations are as follows. To carry out the 
weak-coupling expansion for the theory given by eq.(\ref{e1}) a formal 
procedure based on integrating out the high-frequency components of the 
local spins $S_{i}^{a}$ will be applied. We shall make use of the 
following parametrization of spin variables \cite{a13}

\begin{eqnarray}
S_{k}^{b}&=&\pi_{k}^{b} \qquad b=1,2,...,N-1 \nonumber\\
\pi_{k}^{2}&=&\pi_{k}^{b}\pi_{k}^{b*}\nonumber\\
S^{N}_{k}&=&\sqrt{1-\pi_{k}^{2}} ~ \exp(i\phi_{k})
\label{e5}
\end{eqnarray}

\noindent
where $\pi^{b}_{k}$ are small and slowly-varying fluctuations about the
$N$th-component's direction. Substituting the representation eq.(\ref{e5}) 
into eq.(\ref{e4}) and integrating out the modulus $|S^{N}_{k}|$, one 
arrives at the Lagrangian

\begin{eqnarray}
Z&=&\int \prod_{i}d^{N-1}\pi_{i}d^{N-1}\pi_{i}^{*}
\exp\{-H/T\}\nonumber\\
H&=&H_{0}+H_{int}\nonumber\\
H_{0}&=&-J \sum_{<i,j>}|\pi_{i}^{b}
\exp(iA_{ij})-\pi_{j}^{b}|^{2}\nonumber\\
H_{int}&=&-J\sum_{<i,j>}|\sqrt{1-\pi_{i}^{2}} ~ \exp(\phi_{i}
-\phi_{j}+A_{ij})-
\sqrt{1-\pi_{j}^{2}}|^{2}
\label{e6}
\end{eqnarray}

\noindent
Here the Jacobian factor equals unity. The advantages of this
parametrization are quite evident because if $T\rightarrow 0$ the
spin-wave fields $\pi_{i}^{b}$ can be treated as free lattice fields. As 
these fields fluctuate, a contribution to the effective action for the 
phases $\phi_{i}$ will arise. At the same time, the field $\phi(i)$ does 
not react back onto $\pi_{i}^{b}$ in the renormalization procedure.

The calculations in the one-loop approximation may be readily carried out
by means of the standard momentum-shell recursion relation technique
developed in \cite{a14} (see, also \cite{a13,a15}). To produce a 
systematic low-temperature perturbation theory one has to expand 
nonlinearities such as $\sqrt{1-\pi_{i}^{2}}$ in $H_{int}$ in powers of 
$\pi_{i}^{2}$ and to integrate out short-wavelength degrees of freedom.
Let us decompose the Fourier-transformed spin field according to

\begin{eqnarray}
\pi^{b}(q)&=&\pi_{<}^{b}(q)+\pi_{>}^{b}(q)\nonumber\\
\pi^{b}(q)&=&\pi_{<}^{b}(q) \qquad \qquad 0<q<\Lambda^{\prime}
\nonumber\\
\pi^{b}(q)&=&\pi_{>}^{b}(q) \qquad \qquad \Lambda^{\prime}<q<\Lambda
\label{e7}
\end{eqnarray}

\noindent
with the purpose of integrating out the short-wavelength fields
$\pi_{>}^{b}(q)$. Here $\Lambda$ and $\Lambda^{\prime}$ are momentum
cutoffs and $q$ stands here for the reciprocal lattice coordinates. 
Calculations in the lowest order of perturbation theory yield

\begin{equation}
\frac{H_{int}}{T}=\frac{2}{T}\sum_{<i,j>}\cos(\phi_{i}-\phi_{j}+A_{ij})
(1-\frac{1}{2}<\pi_{i}^{2}>-\frac{1}{2}<\pi_{j}^{2}>)
\label{e8}
\end{equation}

\noindent
where the angular brackets $< \cdots >$ stand for the straight Gaussian 
integration over the modes $\pi_{>}(q)$, and $J$ is absorbed in $T$.
From eq.(\ref{e8}) the temperature renormalization is easily seen to be

\begin{equation}
\frac{1}{T^{\prime}}=\frac{1}{T}(1-<\pi_{i}^{2}>)
\label{e9}
\end{equation}

\noindent
Setting the magnetic field (or, equivalently, the frustration) to zero
one may easily take the long-distance limit of the model and find the bare 
propagator for the $\pi_{>}^{b}(q)$ fields, namely

\begin{equation}
G_{ab}(q)=<\pi_{>}^{a}(q)\pi_{>}^{b}(-q)>=\frac{T}{q^{2}}\delta_{ab}
\label{e10}
\end{equation}

\noindent
The most divergent part of the Goldstone propagator behaves like
$1/q^{2}$ for small $q$, giving rise to the dominant logarithmic term
in $<\pi_{i}^{2}>=(N-1)\ln\frac{\Lambda}{\Lambda^{\prime}}$. Carrying out
the trivial calculations as described in detail in \cite{a13}, one is lead 
to the familiar expression for the one-loop beta function

\begin{equation}
\beta(T)=(d-2)T-\frac{(N-1)T^{2}}{\pi}
\label{e11}
\end{equation}

\noindent
The problem now is to extend this approach to the uniformly frustrated 
model of eq.(\ref{e1}), that is when the magnetic field is switched back 
on.

\subsection{Lattice frustrated models}

Since we are, in this paper, interested in the critical behavior of the 
lattice frustrated model, we shall first of all note some properties of 
the Azbel-Harper-Hofstadter operator $\hat{L}$ \cite{a21,a22,a23,a24}
describing a 2D Bloch charged particle subject to a uniform magnetic field.
Here $\hat{L}$ is the inverse of the quadratic part of the action $H_{0}$ 
which one may formaly regard as the electron hopping Hamiltonian. Using 
a compact notation, the Hamiltonian $H_{0}$ in eq.(\ref{e6}) can be 
rewritten as

\begin{equation}
H_{0}=\sum_{n}\pi_{n}^{b*}\hat{L}\pi_{n}^{b}
\label{e12}
\end{equation}

\noindent
where by definition (due to the choice of the most conventional Landau 
gauge) $\hat{L}$ acts on the Bloch wave function in the following way

\begin{eqnarray}
\hat{L}\pi^{b}(n)&=&\exp(ik_{x})\pi^{b}_{k}(n-1)
+\exp(-ik_{x})\pi^{b}_{k}(n+1)\nonumber\\
&+&2\cos(k_{y}+2\pi\frac{np}{q})\pi^{b}_{k}(n)\nonumber\\
\pi^{b}(n)&=&\exp(ikn)\pi^{b}_{k}(n)
\label{e13}
\end{eqnarray}

\noindent
where $n=1,...,q$ is a coordinate in the magnetic cell and the wave
vector $k$ ranges over the reduced Brillouin zone: $-\pi/a<k_{x}<\pi/a$, 
$-\pi/qa<k_{y}<\pi/qa$. It is worth noting that if $n$ labels the components 
of a $q$-component ``vector'' $\pi^{b}_{k}(n)$, then the operator 
$\hat{L}$ acts as a $q \times q$ Hermitean matrix. In the case of $q=4$ 
one has, for example

\begin{equation}
\hat{L} = \left(
\begin{array}{llll}

2\cos(k_{x})         \qquad & \exp(-ik_{y})        \qquad & 
0                    \qquad & \exp(ik_{y}) \\

\exp(ik_{y})         \qquad & 2\cos(k_{x}+2{\pi}f) \qquad & 
\exp(-ik_{y})        \qquad & 0            \\

0                    \qquad & \exp(ik_{y})         \qquad & 
2\cos(k_{x}+4{\pi}f) \qquad & \exp(-ik_{y}) \\

\exp(-ik_{y})        \qquad & 0                    \qquad &
\exp(ik_{y})         \qquad & 2\cos(k_{x}+6{\pi}f)

\end{array}
\right)
\end{equation}

\noindent
Therefore, this Hamiltonian $H_{0}$ can be thought of as that 
corresponding to a particle hopping along $q$ sites around a ring
\cite{a25,a26}. The spectrum of $H_{0}$ is known to posess a finite set 
of $q$ magnetic sub-bands, each state in these sub-bands being $q$-fold 
degenerate instead of the infinite degenerancy inherent to the 
continuous problem \cite{a26}.

The magnetic translational symmetry properties result from the local 
gauge invariance of the lattice theory, eq.(\ref{e1}). This implies that 
under the translation ${\bf {r}} \rightarrow {\bf{r+a}}$ the propagator 
in eq.(\ref{e10}) transforms as follows

\begin{equation}
G_{ab}({\bf r+a},{\bf r}^{\prime}+{\bf a})
=\exp\{\frac{i\pi}{\Phi_{0}}[{\bf B}\times{\bf a}]({\bf r-r}^{\prime})\}
G_{ab}({\bf r,r}^{\prime})
\label{e14}
\end{equation}

\noindent
From eq.(\ref{e14}) it follows that $G_{ab}(\bf{r,r}^{\prime})$ at
coinciding points does not depend on $\bf{r}$. On the other hand, for the 
propagator from eq.(\ref{e14}) one has the resolvent spectral decomposition

\begin{equation}
G_{ab}({\bf r,r}^{\prime})
=\delta_{ab}\sum_{n=1}^{q}\sum_{\alpha=1}^{q}\sum_{\bf{k}}
E_{n {\bf k}}^{-1}\Psi_{n {\bf k} \alpha}({\bf r})
\Psi_{n {\bf k} \alpha}^{*}({\bf r}^{\prime})
\label{e15}
\end{equation}

\noindent
where $n=1,...,q$ labels the magnetic band number, and 
$\Psi_{n {\bf k} \alpha}({\bf r})$ are eigenfunctions of $\hat{L}$ defined 
in the magnetic Brillouin zone, $-\pi/a<k_{x}<\pi/a$ and 
$ -\pi/qa<k_{y}<\pi/qa$, forming the basis of the $q$-dimensional 
irreducible projective representation of the magnetic translation group.
The exact energy spectrum $E_{n {\bf k}}$ does not depend on the quantum 
number $\alpha=1,...,q$.

There exists a remarkable property of the energy spectrum resulting
from the local gauge invariance of the lattice theory,  eq.(\ref{e1}).
From this local gauge invariance it follows that $E_{n {\bf k}}$ depends 
on ${\bf k}$ only through the parameter 
$\Delta=2-\cos(qk_{x})-\cos(qk_{y})$ and of course on $q$ \cite{a16}

\begin{equation}
E_{n {\bf k}}=E_{n}(\Delta,q)
\label{e16}
\end{equation}

\noindent
From the point of view of the critical properties of the nonlinear
$\sigma$-model under consideration, one can see that only states lying
near the bottom of the lowest magnetic sub-band ($n=1$) are relevant.
Here it works in a similar way, as in the well-known lowest Landau level 
(LLL) projection approximation for the continuous theory.

Within the effective mass approximation the energy spectrum as a
function of a quasimomentum near the band's bottom $(\Delta=0)$ reads

\begin{equation}
E_{1 {\bf k}}=e_{0}(q)+\frac{e_{2}(q)}{2}\{(qk_{x})^{2}+(qk_{y})^{2}\}
\label{e17}
\end{equation}

\noindent
with $e_{0,2}(q)$ being coefficients of the Taylor expansion of 
$E_{1 {\bf k}}$ in powers of ${\bf k}$. Here $e_{2}(q)^{-1}$ is 
proportional to an effective mass in the $(x,y)$-plane, $e_{0}(q)$ being 
proportional to a chemical potential. It is here rather essential that 
the effective mass and chemical potential are some functions of $q$.

In order to find $\beta(T)$ we have to calculate $<\pi_{i}^{2}>$, which 
is given by

\begin{eqnarray}
<\pi_{i}^{2}>&=&\sum_{a=1}^{q}G_{aa}({\bf r},{\bf r}) \nonumber \\
&=&q\sum_{n=1}^{q}\sum_{{\bf k}}E_{n {\bf k}}^{-1}
\label{e18}
\end{eqnarray}

\noindent
The factor $q$ appears as a result of the degeneracy of the spectrum 
and summation over $\alpha$.
Making use the effective mass approximation for the lowest
magnetic sub-band, and evaluating the integral over ${\bf k}$ in 
eq.(\ref{e18}), we arrive at the following expression

\begin{eqnarray}
<\pi_{i}^{2}>&=&q\sum_{{\bf k}}\frac{2e_{2}(q)^{-1}}{(qk_{x})^{2}
+(qk_{y})^{2}+m_{0}^{2}} \nonumber \\
&=&\frac{2(N-1)q}{e_{2}(q)}\ln\frac{\Lambda}{\Lambda^{\prime}}.
\label{e19}
\end{eqnarray}

\noindent
Here, $m_{0}^{2}=2e_{0}(q)/e_2(q)$ denotes some effective mass. With 
the help of eq.(\ref{e9}) and of eq.(\ref{e19}), one can readily derive
the beta-function in the one-loop approximation

\begin{equation}
\beta(T,q)=(d-2)T-\frac{2(N-1)qT^{2}}{\pi e_{2}(q)}
\label{e20}
\end{equation}

\noindent
This expression is the lattice version of eq.(\ref{e11}). It leads to the 
nontrivial ultraviolet-stable fixed point located at

\begin{equation}
T^{*}(q)=\frac{(d-2)\pi e_{2}(q)}{2(N-1)q}
\label{e21}
\end{equation}

\noindent
We see that, in fact, eq.(\ref{e21}) gives the phase transition line
$H_{c2}(T)$ of the lattice model, eq.(\ref{e1}). The peculiarity of
the beta-function, eq.(\ref{e20}), lies in its explicit dependence on the
frustration $q$ revealing a multifractal type structure. This follows from 
the properties of the spectrum of a single-electron in a magnetic field
and a periodic potential (leading to the so-called ``Hofstadter's 
butterfly'' multifractal structure \cite{a22,a23}). From the physical 
point of view this peculiarity looks quite natural since the phase 
transition temperature must depend on the applied magnetic field. After 
setting $q=1$, or $H=0$, $e_{2}(q)$ becomes $e_{2}(1)=2$ and the function 
$\beta(T)$ coincides with eq.(\ref{e9}).

The critical exponent for the correlation length $\nu$ follows immediately
by differentiating eq.(\ref{e20}) at the fixed point. Keeping only the
leading term in $\epsilon$ one obtains

\begin{equation}
\nu=-\frac{1}{\beta^{\prime}(T^{*})}=\frac{1}{\epsilon}
\label{e21b}
\end{equation}

\noindent
To determine Fisher's critical exponent $\eta$ one may apply the 
conventional momentum-shell RG approach to first order in $\epsilon$ as 
described in detail in \cite{a14,a15}. After carrying out the standard 
RG computational procedure, we are led to the following expression

\begin{equation}
\eta(T)=2-d+\frac{(2N-1)qT}{\pi e_{2}(q)}
\label{e22}
\end{equation}

\noindent
Note that eq.(\ref{e21}) and eq.(\ref{e22}) yield the usual one-loop 
result $\eta=\frac{\epsilon}{2(N-1)}$.

We have seen above that the $(2+\epsilon)$-dimensional frustrated lattice 
model undergoes a {\it continuous} phase transition on the coexistance 
curve $H_{c2}(T)$ and is characterized by universal critical exponents 
which we have calculated. A physical interpretation of this conclusion is 
quite simple: the infinite degenerancy inherent to a charged particle 
moving in a uniform magnetic field is lifted by a commensurate potential
suppressing the dimensional reduction effect and favouring the flux lattice 
state \cite{a6}. At least at the lowest order in $\epsilon$ the frustration 
$q$ drops out of the critical exponents, these being identical to those for
the conventional $0(2N)$-symmetric Heisenberg ferromagnet. For $N=1$ the 
expressions obtained above obviously show a singularity in the factor 
$\frac{1}{N-1}$, which reflects the special properties of the neutral
2D XY model.

\section{WEAK-COUPLING EXPANSION FOR THE GINZBURG-LANDAU MODEL}
\renewcommand{\theequation}{3.\arabic{equation}}
\setcounter{equation}{0}

The approach which has been developed until now may be extended to the 
continuous GL model subject to a uniform magnetic field ${\bf B}$ in 
$d=4+\epsilon$ dimensions, in the spirit of the work of Lawrie and Athorne 
\cite{a17,a18}. This model is described by the Hamiltonian

\begin{equation}
H=\int d^{d}x|(\partial_{\mu}+i\frac{2\pi}{\Phi_{0}}
A_{\mu})\Psi_{a}|^{2}
\label{e43}
\end{equation}

\noindent
where the summations over $\mu=1,...,d $ and $a=1,..,N$ are understood in 
eq.(\ref{e43}). Here $\Psi=[\Psi_{1},...,\Psi_{N}]$ is an $N$-component 
complex order parameter, the fixed-length constraint $|\Psi|^{2}=1$ 
being imposed on the local fields. The vector potential $A_{\mu}$ is now
taken within the symmetric gauge

\begin{equation}
{\bf A}=\frac{1}{2} {\bf B}\times{\bf r}
\label{e44}
\end{equation}

\noindent
where ${\bf B}$ is taken along the $z$ axis.
The partition function associated with eq.(\ref{e43}) reads

\begin{equation}
Z=\int \prod_{a=1}^{N}D\Psi_{a}D\Psi_{a}^{*}\exp(-\frac{H}{T})
\delta(|\Psi|^{2}-1)
\label{e45}
\end{equation}

\noindent
We shall make use of the same parametrization of $\Psi_{a}$ exploited
in Section II

\begin{eqnarray}
\Psi_{b}({\bf r})&=&\pi_{b}({\bf r}) \qquad  b=1,2,...,N-1 \nonumber \\
\pi^{2}({\bf r})&=&\pi_{b}({\bf r})\pi_{b}^{*}({\bf r}) \nonumber \\
\Psi_{N}({\bf r})&=&\sqrt{1-\pi^{2}({\bf r})} ~ \exp(i\phi({\bf r}))
\label{e46}
\end{eqnarray}

\noindent
Inserting eq.(\ref{e46}) into eq.(\ref{e45}) and integrating out the
$|\Psi_{N}|$, we obtain the effective Hamiltonian convenient for 
generating the low-temperature expansion

\begin{equation}
H=\int d^{d}x[|(\partial_{\mu}
+i\frac{2\pi}{\Phi_{0}}A_{\mu})\pi_{b}|^{2}+
(\partial_{\mu}\phi +\frac{2\pi}{\Phi_{0}}A_{\mu})^{2}
+(\partial_{\mu}\sqrt{1-\pi^{2}})^{2}]
\label{e47}
\end{equation}

\noindent
Observe that no Jacobian arises from the elimination of $|\Psi_{N}|$.
To determine the temperature renormalization in the lowest order of
perturbation theory, we have to take into account only the first two
relevant terms in eq.(\ref{e46}) and disregard the third one
describing the interaction between the transverse degrees of freedom 
$\pi_{b}$. The RG equation obtained in this way is

\begin{equation}
\frac{1}{T^{\prime}}=\frac{1}{T}(1-<\pi^{2}({\bf r})>_{s})
\label{e48}
\end{equation}

\noindent
where $< \cdots >_{s}$ stands for integrating over the short-wavelength
transverse fields $\pi_{b}$.

Notice that the non gauge-invariant correlation function

\begin{equation}
G_{ab}({\bf r},{\bf r}^{\prime})
=<\pi_{a}({\bf r})\pi_{b}({\bf r}^{\prime})>
\label{e49}
\end{equation}

\noindent
is the Green function of the $d$-dimensional Schrodinger operator

\begin{equation}
(-i\partial_{\mu}-\frac{2\pi}{\Phi_{0}}A_{\mu})^{2}
G_{ab}({\bf r},{\bf r}^{\prime})=\delta_{ab}\delta({\bf r}
-{\bf r}^{\prime})
\label{e50}
\end{equation}

\noindent
The exact solution of eq.(\ref{e50}) in an arbitrary gauge reads \cite{a20}

\begin{eqnarray}
G_{ab}({\bf r},{\bf r}^{\prime})&=&\delta_{ab}
\exp \left ( -i\frac{2\pi}{\Phi_{0}}
\int_{{\bf r}}^{{\bf r}^{\prime}}dx_{\mu}A_{\mu} \right )
(4\pi)^{\frac{2-d}{2}}\int_{0}^{\infty}du
\frac{u^{\frac{2-d}{2}}}{2\sinh(u\omega/2)} \nonumber \\
&\times& \exp\{-\frac{(z-z^{\prime})^{2}}{4u}-\frac{\omega}{8}\coth
[\frac{1}{2}u\omega((x-x^{\prime})^{2}+(y-y^{\prime})^{2})]\}
\label{e51}
\end{eqnarray}

\noindent
where $\omega=\frac{2eB}{c}$ is the cyclotron frequency (here we have set 
$\hbar=1$ and $2m=1$) and $z$ and $z^{\prime}$ are $(d-2)$-dimensional 
longitudinal coordinates. The integral in eq.(\ref{e51}) is taken over 
the straight line connecting the points ${\bf r}$ and ${\bf r}^{\prime}$.

Near the phase boundary $H_{c2}(T)$, only the lowest Landau level gives a
dominant contribution to the gauge-invariant quantity $G({\bf r},{\bf r})$
entering eq.(\ref{e48}). The LLL approximation leads to a nice 
simplification of eq.(\ref{e51})

\begin{eqnarray}
G_{ab}({\bf r},{\bf r}^{\prime})&=&\delta_{ab}
\exp\{-i\frac{2\pi}{\Phi_{0}}
\int_{{\bf r}}^{{\bf r}^{\prime}}dx_{\mu}A_{\mu}
-\frac{\omega}{8}[(x-x^{\prime})^{2}+(y-y^{\prime})^{2}]\} \nonumber \\
&\times&(4\pi)^{-\frac{d}{2}}\int_{0}^{\infty}du
u^{\frac{2-d}{2}}\exp\{-\frac{u\omega}{2}
-\frac{(z-z^{\prime})^{2}}{4u}\}
\label{e52}
\end{eqnarray}

\noindent
In going from eq.(\ref{e51}) to eq.(\ref{e52})  we have made the following 
substitution in eq.(\ref{e51}): $\coth(u\omega/2) \rightarrow 1$ and
$\sinh(u\omega/2) \rightarrow \frac{1}{2}\exp(u\omega/2)$ (justified by
the limit $B \rightarrow \infty$ of the LLL approximation).

To make further calculations more transparent, we will make use of the 
mixed coordinate-momentum representation for
$G_{ab}({\bf r},{\bf r}^{\prime})$. After carrying out the Fourier 
transformation in the longitudinal variables $z, z^{\prime}$ eq.(\ref{e52}) 
becomes

\begin{equation}
G_{ab}(x,x^{\prime},y,y^{\prime},{\bf k})
=\delta_{ab}\exp\{-i\frac{2\pi}{\Phi_{0}}
\int_{{\bf r}}^{{\bf r}^{\prime}}dx_{\mu}A_{\mu}
-\frac{\omega}{8}[(x-x^{\prime})^{2}+(y-y^{\prime})^{2}]\}
\frac{\omega}{k^{2}+\xi^{2}}
\label{e53}
\end{equation}

\noindent
where ${\bf k}$ is a $d-2$-dimensional vector and $\xi$ is a correlation
length in the longitudinal directions. The important conclusion one may
draw from eq.(\ref{e53}) is that a dimensional reduction effect takes
place in the physics of our model, eq.(\ref{e43}). We have also set an 
irrelevant factor to unity.

With the result of eq.(\ref{e53}) at hand, we can now readily evaluate
$<\pi^{2}({\bf r})>_{s}$

\begin{equation}
<\pi^{2}({\bf r})>_{s}=\frac{(N-1)\omega}{\pi}
\ln\frac{\Lambda}{\Lambda^{\prime}}
\label{e54}
\end{equation}

\noindent
Eq.(\ref{e48}) and eq.(\ref{e54}) yield the beta-function in the one-loop
approximation as well as the phase boundary, namely

\begin{eqnarray}
\beta(T,B)&=&(d-4)T-\frac{(N-1)\omega}{\pi}T^{2}\nonumber\\
T^{*}(B)&=&\frac{\pi(d-4)}{(N-1)\omega}
\label{e55}
\end{eqnarray}

\noindent
It is remarkable that the calculation of the critical exponents, like the
one carried out in Section II, leads now to the same universal values as
given by eq.(\ref{e21}) and eq.(\ref{e22}), independently of $B$.

We have seen that in contrast to the results obtained e.g. in \cite{a3} for 
the $(6-\epsilon)$-dimensional case, a {\it continuous} phase transition 
occurs in the GL model in $4+\epsilon$ dimensions. It is described by the 
critical exponents of the $0(2N)$-symmetric Heisenberg ferromagnet (in the 
one-loop approximation). From the field-theoretical point of view we are 
now dealing with the dimensional reduction effect within the Abelian Higgs 
model defined by an Euclidean Lagrangian, eq.(\ref{e43}) (or, equivalently,
in the $N$-component scalar QED), in a large external magnetic field. The 
analogous effect was recently shown to occur also in the conventional $4D$ 
spinor QED subject to an extremely high magnetic field like in a vicinity 
of a neutron star \cite{a19}.

\section{Conclusions}
\renewcommand{\theequation}{6.\arabic{equation}}
\setcounter{equation}{0}

It has been shown that the $(2+\epsilon)$-dimensional $SU(N)$-symmetric
uniformly-frustrated lattice spin model undergoes a second-order phase 
transition, described by the universal critical exponents of the 
$0(2N)$-symmetric Heisenberg ferromagnet irrespectively of the value of the 
frustration. Our calculations provide no evidence of a first-order phase 
transition. This result was found to hold both for the 
$(2+\epsilon)$-dimensional lattice model as well as for the 
$(4+\epsilon)$-dimensional continuos GL model in an applied magnetic field. 
The occurence of the second-order phase transition is in this case in 
contrast to the RG calculations carried out in $(6-\epsilon)$-dimensions 
\cite{a3}. Contrary to the conventional RG approach based on the 
$\phi^{4}$-theory \cite{a9}, the low-temperature RG approach employed in 
this paper works for an arbitrary rational value of the frustration 
$f=\Phi/\Phi_{0}$. It allows us to compute the phase transition boundary 
line $H_{c2}(T)$, which is a very difficult calculation problem within the 
standard $\phi^{4}$-theory. In the case of the lattice model the 
coexistence curve exhibits a very complicated structure (in fact, 
reminiscent of the ``devil's staircase'' \cite{a9} inherited from the 
Hofstadter's butterfly's structure \cite{a22,a23}) which can be determined 
only numerically for all integer values of $q$.

Our treatment can be easily extended to the large-$N$ limit.
But the situation here is somewhat unclear since different groups of 
researchers have come to different conclusions concerning the nature of 
the transition. On the one hand, our results -- being in good agreement 
with those obtained within the $\frac{1}{N}$-expansion by Radzihovsky 
\cite{a6} and by the authors of \cite{a5} -- give evidence in favour of a 
continuous phase transition. In particular, the non-trivial fixed point 
responsible for the second-order phase transition was found. Moreover, 
from the equation for the correlation length $\xi$ it follows that there 
is a divergence of $\xi$ at crticality. One the other hand, the 
calculation of the effective potential within the $\frac{1}{N}$-expansion 
as presented in \cite{a4}, gives evidence for a first-order phase 
transition. The origin of this discrepancy in conclusions is rather subtle 
and probably lies in non-commutativity of the large-$N$ limit and the 
thermodynamic limit (see, for a discussion, ref. \cite{a6}).

The main problem still open for future research is how to extend the 
results obtained by means of the low-temperature RG approach to the 
$N=1$ case. Thus, we have seen that the underlying lattice plays an 
essential important role near the upper critical magnetic field $H_{c2}(T)$
restoring (like weak disorder) the continuos phase transition into the 
flux-lattice state. This is indeed the main result of our paper and we 
have commented in the Introduction on possible direct applications of the
model we have studied.

\section{Acknowledgements}

This work was supported by the Russian Foundation for Basic Research
Grant No. 96-02-16858, the NATO Collaborative Research Grant No.
OUTR.CRG960838 and by the EC contract No. ERB4001GT957255. The authors
are most grateful to the Max-Planck-Institute f{\"u}r Physik komplexer 
Systeme, Au{\ss}enstelle Stuttgart, where a considerable part of this work 
was carried out, for kind hospitality and the use of its facilities. One 
of the authors (BNS) is also most grateful to the Department of High 
Energy Physics of the International School for Advanced Studies in 
Trieste, where part of this work was carried out, for support and warm 
hospitality. He is deeply grateful to S.Shenoy and R.Iengo for 
interesting discussions and valuble comments. It is also a great pleasure 
to thank H.Capellmann, G.Eliashberg, A.Nersesyan, A.Protogenov and W.Selke 
for helpful discussions.

\end{document}